\documentclass[
   ,final            % use final for the camera ready runs
  ]
  {aipproc}
\layoutstyle{6x9}

\usepackage{graphicx}
\usepackage{colordvi}

    \def\be{\begin{equation}}
    \def\ee{\end{equation}}
    \def\ea{\end{eqnarray}}

\begin{document}

\title{Stochastic growth of quantum fluctuations during inflation}

\classification{04.62.+v, 98.80.Cq}
\keywords      {Inflation; Stochastic Approach; Gauge Invariant Variables}

\author{F. Finelli}{
address={INAF/IASF Bologna, Istituto di Astrofisica Spaziale e Fisica Cosmica di Bologna 
via Gobetti 101, I-40129 Bologna, Italy and\\
INFN, Sezione di Bologna, Via Irnerio 46, I-40126 Bologna, Italy}}
%,altaddress={E-mail: finelli@iasfbo.inaf.it}

\author{G. Marozzi}{
address={Coll\`ege de France, 11 Place M. Berthelot, 75005 Paris, France and \\
GR$\varepsilon$CO --
Institut d'Astrophysique de Paris, UMR7095, CNRS,  \\
Universit\'e Pierre \& Marie Curie, 98 bis boulevard Arago,
75014 Paris, France}}

\author{A. A. Starobinsky}{
address={Landau Institute for Theoretical Physics, Moscow, 119334, Russia and \\
Research Center for the Early Universe, Graduate School of Science,
The University of Tokyo, Tokyo 113-0033, Japan}}

\author{G. P. Vacca}{
address={INFN, Sezione di Bologna, Via Irnerio 46, I-40126 Bologna, Italy and \\
Dipartimento di Fisica, Universit\`a degli Studi di
Bologna, via Irnerio 46, I-40126 Bologna, Italy}}

\author{G. Venturi}{
address={INFN, Sezione di Bologna, Via Irnerio 46, I-40126 Bologna, Italy and \\
Dipartimento di Fisica, Universit\`a degli Studi di
Bologna, via Irnerio 46, I-40126 Bologna, Italy}}

%%%%%%%%%%%%%%%%%%%%%%%%%%%%%%%%%%%%%%%%%%%%%%%%%
\begin{abstract}
The standard field-theoretical approach to the 
slow-roll inflation is introduced. We then show as, in order to  
calculate the mean square of the canonical gauge invariant quantum fluctuations  
associated to a generic field, the logarithm of the scale factor has  
to be used as the time variable in the Fokker-Planck equation in the  
stochastic approach.
Then we compute the growth of different test fields with a small
effective mass during slow-roll inflationary models,  
comparing the results with the one for the gauge invariant canonical fluctuation 
associated to the inflaton, the Mukhanov variable. 
We find that in most of the single fields inflationary  
models such fluctuation grows  
faster than any test field with a non-negative effective mass,
with the exception of hybrid models.
\end{abstract}

\maketitle

%%%%%%%%%%%%%%%%%%%%%%%%%%%%%%%%%%%%%%%%%%%%
\section{I. Introduction}
%%%%%%%%%%%%%%%%%%%%%%%%%%%%%%%%%%%%%%%%%%%%%%%%%%%%%%

The theory of quantum fields in an expanding universe has evolved
from its pioneering years \cite{BD} into a necessary tool in order
to describe the Universe on large scales. The de Sitter background
- characterized by the Hubble parameter $H\equiv \dot a/a$ being
constant in time (for the flat spatial slice), where $a(t)$ is the
scale factor of a Friedmann-Lemaitre-Robertson-Walker (FLRW) cosmological
model - has been the main arena to compute quantum effects
even before becoming a pillar of our understanding of the
early inflationary stage and of the recent acceleration of the
Universe.

However, while $|\dot H|\ll H^2$ for any inflationary model, $\dot
H$ may not become zero in a viable model, apart from some isolated
moments of time. Indeed, the standard slow-roll expression for the
power spectrum of the adiabatic mode of primordial scalar
(density) perturbations becomes infinite, i.e., meaningless, if
$\dot H$ becomes zero during inflation.
Therefore, the study of quantum effects in a nearly
de Sitter stage with $\dot H \not= 0$, in particular, when the
total change in $H$ during inflation is not small compared with
its value during the last e-folds of inflation
\cite{FMVV_I,FMVV_II,FMVV_IV,us1, us2}
(see also the recent papers
\cite{UM,JMPW,JP}), is not of just pure theoretical interest.

In general inflationary model quantum scalar field can be split into a long 
wave (coarse grained) component and a short-wave (perturbative) one.
Then it can be proved that the former component effectively becomes
quasi-classical, though random (i.e. all non-commutative parts of it
may be neglected), and it experiences a random walk described by 
the stochastic inflation approach \cite{S86}.

However, it should be emphasized that just because of the non-perturbative
nature of the stochastic approach to inflation, it is based on a number 
of heuristic approximations. Therefore, it is very important to check,
whenever possible, results obtained by its application using the standard
perturbative QFT in curved space-time.

In this paper we discuss, following the result of \cite{us1, us2}, the diffusion equation for 
general scalar fluctuations in a generic model of inflation.
On using the results obtained by field theory methods,
we show that the stochastic diffusion equation for the canonical gauge invariant variables
associated with these generic scalar fluctuations should be
formulated in terms of the number of e-folds $N$.

We then tackle in more detail the moduli problem
issue. Following \cite{us1,us2} we consider the stochastic growth of different test 
fields as massive minimally coupled scalar fields,
massless non-minimally coupled scalar fields and moduli
with an effective mass $\propto H^2$, in 
different inflationary models.
On the other hand, it is known 
\cite{FMVV_II} that the mean square of
gauge invariant Mukhanov variable \cite{Mukhanov1988} grows; it is therefore interesting to
compare the amplification of the test fields above not only with the
background inflaton dynamics, but also with the stochastic growth
of this gauge invariant fluctuation. This comparison aims for
a self-consistent understanding of quantum foam during inflation.

The paper is organized as follows.
In Section II we describe a two field model with two generic 
self-interacting potentials,
each of them depends on one field only,
introducing the field theory approach.
In Section III we examine the stochastic approach for such two field model.
In Section IV we derive the equations to obtain the stochastic growth of 
the test fields considered and of the gauge invariant Mukhanov variable.
In Section V we consider four representative cases of the inflationary 
{\em ``zoo''} for which we calculate the growth of the fields and compare different results.
In Section VI we present our conclusions.

%%%%%%%%%%%%%%%%%%%%%%%%%%%%%%%%%%%%%%%%%%%%%%%%%%%%%%%%%%%%%%%%%%%%%%%%%%%%%%%%%%%%

\section{II. Inflation and field-theoretical approach}

Let us consider, in a spatially flat 
FLRW background geometry, a  two
 field model in which the dynamics is driven by a minimally coupled inflaton
$\phi$ and a minimally coupled scalar field $\chi$ is present.
%(see \cite{LMY} for a different approach to the moduli problem).
We shall neglect the $\chi$ energy density and pressure in the
background FLRW equations.

The action is given by 
\begin{eqnarray}
    S  &=& \int d^4x \sqrt{-g} \left[ 
\frac{R}{16{\pi}G}
    - \frac{1}{2} g^{\mu \nu}
    \partial_{\mu} \phi \partial_{\nu} \phi
    - V(\phi) 
- \frac{1}{2} g^{\mu \nu}
    \partial_{\mu} \chi \partial_{\nu} \chi
    - \bar{V}(\chi) \right] 
    \label{action}
\end{eqnarray}

and we can expand our background fields $\{\phi$, $\chi$, $g_{\mu\nu}\}$ up 
to second order in the non-homogeneous perturbations, without fixing any gauge, as 
follows:
\be
\phi(t,\vec{x})=\phi^{(0)}(t)+\phi^{(1)}(t,\vec{x})+\phi^{(2)}(t,\vec{x})
\,\,\,\,\,,\,\,\,\,\,
\chi(t,\vec{x})=\chi^{(0)}(t)+\chi^{(1)}(t,\vec{x})+\chi^{(2)}(t,\vec{x})\,,
\ee
\begin{eqnarray}
& & g_{00}= -1-2 \alpha^{(1)}-2 \alpha^{(2)}, ~~~~~~~\,\,\,\,\,\,\,\,\,\,
g_{i0}=-{a\over2}\left(\beta^{(1)}_{,i}+\beta^{(2)}_{,i}\right) 
%-{a\over2}\left(\beta^{(2)}_{,i}+B^{(2)}_i\right)
\,,
\nonumber \\
& & 
g_{ij} = a^2 \left[ \delta_{ij} \left(1-2 \psi^{(1)}-2 \psi^{(2)}\right)
+D_{ij} (E^{(1)}+E^{(2)}) \right],
%+ {1\over 2} \left(\chi^{(1)}_{i,j}+\chi^{(1)}_{j,i}+h^{(1)}_{ij}\right) \right.
%\nonumber \\
%& & \left.
%\,\,\,\,\,\,\,\,\,\,\,\,\,\,+
%{1\over 2} \left(\chi^{(2)}_{i,j}+\chi^{(2)}_{j,i}+h^{(2)}_{ij}\right)\right],
\nonumber
\end{eqnarray}

where $D_{ij}=\partial_i \partial_j- \delta_{ij} (\nabla^2/3)$, and we neglect vector 
and tensor perturbations.

The degrees of freedom above are in part redundant.
To obtain a set of well defined equations (Einstein equations + equations 
of motion of the matter sector),  order by order, we have, for example,
to set to zero two modes of scalar perturbations.
The choice of such variables is called a choice of "gauge".

In this case, the study of the scalar sector
can be reduced to the study of two different gauge invariant variables.
Let us choice, as one of these, the so-called Mukhanov variable $Q$
\cite{Mukhanov1988}, which is usually used to canonically quantize the Einstein-Klein-Gordon 
Lagrangian.
The Mukhanov variable can be seen, order by order, as the scalar field fluctuations $\phi^{(n)}$ on uniform curvature hypersurface (see, for example, \cite{Malik2005}).  
To first order, one obtains
\be
Q^{(1)}=\phi^{(1)}+\frac{\dot{\phi}^{(0)}}{H}\left(\psi^{(1)}+\frac{1}{6}
\nabla^2 E^{(1)}\right)\,.
\ee
In the same way we can define, order by order, the second gauge invariant variable as 
the scalar field fluctuations $\chi$
on uniform curvature hypersurface.
We will call this $Q_{\chi}$ and, to first order, it is given by
\be
Q_{\chi}^{(1)}=\chi^{(1)}+\frac{\dot{\chi}^{(0)}}{H}\left(\psi^{(1)}+\frac{1}{6}
\nabla^2 E^{(1)}\right)\,.
\ee
A particular choice of gauge is the one that fix to 
zero $\psi$ and $E$ to all orders. Such a gauge is called the Uniform Curvature 
Gauge (UCG) and in this one obtains
\be
Q^{(n)}=\phi^{(n)} \,\,\,\,\,\,,\,\,\,\,\,\, Q_\chi^{(n)}=\chi^{(n)}
\ee
In the UCG, at the leading order
in the slow-roll approximation and in the long-wavelength limit, 
the equation of motions of the scalar fields $\phi$ and $\chi$
can be obtained, order by 
order, from the expansion around the classical solution of
\be
\frac{d \phi}{d N} = -\frac{V_\phi}{3 \, H(\phi)^2} \, \, \, \, 
\,\, \, \, \, \,, 
\, \, \, \, \,\, \, \, \, \,
\frac{d \chi}{d N} = -\frac{\bar{V}_\chi}{3 \, H(\phi)^2} \,,
\ee
with $N=\log(a(t)/a(t_i))$ being the number of e-folds.

So for the inflaton field we have
\begin{eqnarray}
& & \ddot{\phi}^{(0)}+3 H \dot{\phi}^{(0)}+V_{\phi}=0 \, \, \, \, \,
\, \, \, \, \,, 
\, \, \, \, \,\, \, \, \, \, 3 H \dot{\phi}^{(1)}+\left[V_{\phi\phi}-
\frac{V_{\phi}^2}{3 H^2 M_{pl}^2}\right] \phi^{(1)}=0\,,
\nonumber \\
& & 3 H \dot{\phi}^{(2)}+\left[V_{\phi\phi}-
\frac{V_{\phi}^2}{3 H^2 M_{pl}^2}\right] \phi^{(2)}=-\frac{1}{2}
\left[V_{\phi\phi\phi}-
\frac{V_{\phi\phi}V_{\phi}}{H^2 M_{\rm pl}^2 } + 
\frac{2 V_{\phi}^3}{9 H^4 M_{\rm pl}^4} \right] \phi^{(1)\,2} \,,\nonumber
\end{eqnarray}
while for the field $\chi$ one obtains
\begin{eqnarray}
& & \ddot{\chi}^{(0)}+3 H
\dot{\chi}^{(0)}+\bar{V}_\chi=0
\, \, \, \, \,\, \, \, \, \,, 
\, \, \, \, \,\, \, \, \, \, 3 H \dot{\chi}^{(1)}
+\bar{V}_{\chi\chi}\chi^{(1)}=2 \frac{H_\phi}{H} \bar{V}_\chi \varphi^{(1)}
\nonumber \\
& &  \!\!\!\!\!\!\!\!\!\!\!\!\!\!\!\!\!\!\!\!\!\!\!\!\!3 H \dot{\chi}^{(2)}\!
+\!\bar{V}_{\chi\chi}\chi^{(2)}\!= \! 
2\frac{H_\phi}{H} \bar{V}_\chi \varphi^{(2)}\!
+\!\left[\frac{H_{\phi\phi}}{H}
\!-\!3 \left(\!\frac{H_\phi}{H}\right)^2\right]
\!\bar{V}_\chi \varphi^{(1) 2}\! +\!2 \bar{V}_{\chi\chi}  
\frac{H_\phi}{H} \varphi^{(1)}
\chi^{(1)}
\!-\!\frac{\bar{V}_{\chi\chi\chi}}{2} \chi^{(1) 2}
\nonumber
\end{eqnarray}

\section{III. Stochastic growth of quantum fluctuation}

The results in the previous section suggest that one has to choose the time variable 
$N=\int H(t)dt$ in the Langevin stochastic equation for the large-scale 
part of $\phi$ or $\chi$.

Following \cite{S86}, the number of e-folds $N$ was considered as a time 
variable in the stochastic Langevin equation in a number of papers, e.g. 
in \cite{Gangui} and most recently in \cite{ENPR08}, while in many other 
ones the proper time $t$ was used, e.g. in \cite{LLM,MM05,GT05,
MartinMusso}. This usage of different time variables should not be mixed 
with the invariance of all physical results with respect to a 
(deterministic) time reparametrization $t\to f(t)$ which is trivially 
satisfied after taking the corresponding change in the metric lapse 
function into account. In contrast, the transformation from $t$ to $N$ made 
using the stochastic function $H(\phi(t))$ leads to a physically 
different stochastic process with another probability distribution.   
The new statement made in \cite{us1,us2} is 
that one {\em should} use the $N$ variable when calculating mean squares
of any gauge invariant quantity containing metric fluctuations like the Mukhanov 
variable $Q$ or the gauge invariant variable $Q_{\chi}$. 
Otherwise, incorrect results would be obtained using the 
stochastic approach which would then not coincide with those obtained using 
perturbative QFT methods.

The Langevin stochastic equations 
can so be written as
\begin{eqnarray}
& & \frac{d \phi}{d N} = -\frac{V_\phi}{3H^2} + \frac{f_\phi}{H} \,,
 \, \, \, \, \,\, \, \, \, \,\, \, \, \, \,\, \, \, \, \,\, \, \, \, \, 
\,\, \, \, \, \,\, \, \, \, \,\, \, \, \, \, \,\, \, \, \, \,\, \, \,\, \, \,
\frac{d \chi}{d N} = -\frac{\bar{V}_\chi}{3H^2} + 
\frac{f_\chi}{H}\,,
\nonumber \\ 
& &
\langle f_\phi(N_1) f_\phi(N_2) \rangle =
\frac{H^4}{4\pi^2}\ \delta(N_1 -N_2) \,, \, \, \, \, \,\, \, \,
\langle f_\chi(N_1) f_\chi(N_2) \rangle  =  
\frac{H^4}{4\pi^2}\ \delta(N_1 -N_2)\,,
\nonumber
\end{eqnarray}
where $H^2=V(\phi(t))/3M_{pl}^2$ is a function of $\phi$.
The stochastic noise terms are given, to the leading order in the
slow-roll approximation ($\epsilon=\frac{M_{pl}^2}{2} \left(\frac{V_{\phi}}{V}\right)^2$), by 
\begin{eqnarray} 
f_\phi(t, {\bf x})&=&\epsilon a
H^2 \int \frac{d^3 k}{(2 \pi)^{3/2}} \delta (k-\epsilon a H) \left
[ \hat{a}_k \phi_k(t) e^{- i {\bf k \cdot x}} 
 + \hat{a}^\dagger_k \phi^*_k(t) e^{+i {\bf k \cdot x}}\right]\,,
\nonumber \\
f_\chi(t, {\bf x})&=&\epsilon a
H^2 \int \frac{d^3 k}{(2 \pi)^{3/2}} \delta (k-\epsilon a H) \left
[ \hat{b}_k \chi_k(t) e^{- i {\bf k \cdot x}} 
 + \hat{b}^\dagger_k \chi^*_k(t) e^{+i {\bf k \cdot x}}
\right] \,.
\nonumber 
\end{eqnarray}
On expanding to first order one obtains for the inflaton fluctuation \cite{us1}
\be
\frac{d}{d N}\phi^{(1)}+2 M_{pl}^2\left(\frac{H_{\phi\phi}}{H}
-\frac{H_{\phi}^2}{H^2}\right)\phi^{(1)}=\frac{f_\phi}{H}\,,
\ee
which gives the following result for the growth of quantum fluctuations 
\be
\langle (\phi^{(1)})^2 
\rangle=\frac{1}{4 \pi^2} \left(\frac{V_\phi}{V}\right)^2
\int_{t_i}^t dt' H^3 \left(\frac{V}{V_\phi}
\right)^2\,.
\ee
In the same way, to second order,  we can obtain \cite{us1}
\be
\langle \phi^{(2)} \rangle \!\!=\!\!\left(\frac{V_\phi}{V}
\right)\!\! \int_{t_i}^t\!\! dt' \!\!\left(\frac{V}{V_\phi}
\right)\!\!\left\{ \!\!\frac{H^3}{16 \pi^2} 
\left(\frac{V_\phi}{V}
\right)\!\!+\!\!\frac{1}{2}
\left[ \!\!-\frac{1}{3H} V_{\phi\phi\phi}\!\!+\!\!
\left( \frac{1}{H} V_{\phi\phi} +4
  \frac{\dot{H}}{H}\right) \frac{V_\phi}{V}
\right]\!\!\langle ( \phi^{(1)})^2 
\rangle\right\}\,.
\ee
On the other hand for the growth of quantum $\chi$ fluctuations one
obtains \cite{us2}
\be
\langle \chi^{(1) 2}\rangle=
\frac{\bar{V}_\chi^2}{4 \pi^2} \int^t_{t_i} d\tau \left[
\frac{H(\tau)^3}{\bar{V}_\chi(\tau)^2}-\frac{4}{9 M_{pl}^2} \int^\tau_{t_i}
d\eta \frac{\dot{H}(\tau)}{H(\tau)^3}\frac{\dot{H}(\eta)}{H(\eta)^3}
\int^\eta_{t_i} d\sigma \frac{H(\sigma)^5}{\dot{H}(\sigma)}\right]
\label{chiSquare}
\ee
\be
\langle \chi^{(2)}\rangle \!\!=\!\! \bar{V}_\chi \!\!\int^t_{t_i}\!\!
 d\tau\!\!\left[
\frac{2}{3}\frac{H_\phi}{H^2} \langle \varphi^{(2)}\rangle
\!\!-\!\!\frac{1}{6 H} \frac{\bar{V}_{\chi\chi\chi}}{\bar{V}_\chi} \langle
\chi^{(1) 2}\rangle\!\!+\!\!\frac{2}{3}
\frac{H_\phi}{H^2} \frac{\bar{V}_{\chi\chi}}{\bar{V}_\chi}
\langle \varphi^{(1)} \chi^{(1)}\rangle\!\!+\!\!\frac{1}{3}
\left(\!\!\frac{H_{\phi\phi}}{H^2}\!\!-\!\!3
\frac{H_{\phi}^2}{H^3}\right)\!\! \langle \varphi^{(1) 2}\rangle \right]
\ee
where
\be
\langle \varphi^{(1)} \chi^{(1)}\rangle=-\frac{\bar{V}_\chi}{12 \pi^2}
\frac{\dot{\phi}}{H M_{pl}^2} \int^t_{t_i} d\tau \int^t_{\tau} d\eta
\left[\frac{H(\tau)^5}{\dot{H}(\tau)} \frac{\dot{H}(\eta)}{H(\eta)^3}\right]
\ee

%%%%%%%%%%%%%%%%%%%%%%%%%%%%%%%%%%%%%%%%%%%%%%%%%%%%%%%%%%%%%%%%%%%%%%

\section{IV. Test fields and Mukhanov variable}

We shall consider now 
three different test scalar fields, with a small effective mass and a zero homogeneous
expectation value, and the canonical gauge invariant inflaton fluctuation described by the Mukhanov variable.
In this particular limit the first order fluctuations $\chi^{(1)}$
will be always gauge invariant and coincident with $Q_{\chi}^{(1)}$, independently from the gauge chosen. 

The stochastic growth of $\chi^{(1)}$ can be obtained 
from Eq.(\ref{chiSquare}) in the limit $\chi^{(0)}=0$.
In particular $\langle \chi^{(1)\,2} \rangle$ and $\langle Q^{(1)\,2} \rangle$ 
will be obtained under the natural
assumption of the absence of particles in the in-vacuum state,
more exactly that each Fourier mode ${\bf k}$ of the quantum fields
were in the adiabatic vacuum state deep inside the Hubble
radius and long before the first Hubble radius crossing during
inflation.

%%%%%%%%%%%%%%%%%%%%%%%%%%%%%%%%%%%%%%%%%%%%%%%%%%%%%%%%%%%%%%%%%%%%%%

\subsection{Test scalar field with a constant mass $m_\chi$}

The stochastic equation is
\be
\frac{d \langle \chi^{(1)\,2} \rangle}{d N}
+ \frac{2 m^2_\chi}{3 H (N)^2} \langle \chi^{(1)\,2} \rangle =
\frac{H^2 (N)}{4 \pi^2} \,.
\label{Stoch1}
\ee
Its general solution is
\be
\langle \chi^{(1)\,2} \rangle =
\left(  \int^N_{N_i} \!\!\!dn \frac{H^2(n)}{4 \pi^2} 
e^{\int^n_{N_i} \frac{2 m^2_\chi}{3
 H^2 ({\tilde n})} d{\tilde n}}
\right) e^{-\int^N_{N_i} \frac{2 m^2_\chi}{3 H^2 (n)} dn} \,,
\label{mconstant}
\ee
where we have assumed  $\langle \chi^{(1)\,2} \rangle (N_i)=0$ (we shall
adopt the same choice afterwards).

\subsection{Test scalar field with $m_\chi^2=c H^2$}

If $|c|\ll 1$ the stochastic equation is:
\be
\frac{d
\langle \chi^{(1)\,2} \rangle}{d N} + \frac{2 c}{3}
\langle \chi^{(1)\,2} \rangle = \frac{H^2 (N)}{4 \pi^2} \,.
\ee
Its general solution is
\be
\langle \chi^{(1)\,2} \rangle =
\left(  \int^N_{N_i} d n \frac{H^2(n)}{4 \pi^2} e^{\frac{2}{3} c n}
\right) e^{-\frac{2}{3} c N} \,.
\label{SolB}
\ee

\subsection{Massless non-minimally coupled test scalar field}

The stochastic equation is:
\be
\frac{d \langle \chi^{(1)\,2}
\rangle}{d N} + 4 \xi (2 - \epsilon) \langle
\chi^{(1)\,2} \rangle = \frac{H^2 (N)}{4 \pi^2} \,,
\ee
where $\xi$ is the non-minimal coupling
to the Ricci scalar $R$ and we assume that $|\xi|\ll 1$ (however,
$\xi N$ may be large). 

The term in the action proportional
to $\xi \chi^2 R$ gives an effective time dependent mass $m_\chi^2=6\xi H^2 (2-\epsilon)$
where $\epsilon=-\dot{H}/H^2$.

Its general solution is
\be
\langle \chi^{(1)\,2} \rangle =\left(
\int^N_{N_i} d n \frac{H^{2+4 \xi} (n)}{4 \pi^2 H_i^{4 \xi}} e^{8 \xi
 n}\right)\left( \frac{H_i}{H(N)} \right)^{4 \xi}  e^{- 8 \xi N}
 \,.
\label{SolC}
\ee

\subsection{Mukhanov variable}
%%%%%%%%%%%%%%%%%%%%%%%%%%%%%%%%%%%%%%%%%%%%%%%%%%%%%%%%%%%%%%%%%%%%%

The results for the test fields should be compared 
with the growth of the canonical gauge invariant inflaton fluctuation, namely the Mukhanov variable, $Q^{(1)}$.
The evolution equation for
$\langle Q^ {(1)\,2} \rangle$ found in
\cite{us1} can be re-written as \cite{us2}:
\be
\frac{d \langle Q^ {(1)\,2} \rangle_{\rm REN}}{d N}
+ 2 \left( \eta - 2 \epsilon \right)
\langle Q^ {(1)\,2} \rangle_{\rm REN} =
\frac{H^2 (t)}{4 \pi^2} \,,
\label{inflaton}
\ee
where $\eta = M_{\rm pl}^2 \frac{V_{\phi \phi}}{V}$.

In Eq. (\ref{inflaton}) the positivity of $\eta - 2 \epsilon$ is
not determined by the convexity of the potential,
 i.e. $V_{\phi \phi} > 0$, as we would expect in the absence of
metric perturbations. The threshold corresponds to the 
condition $\frac{d}{d \phi} \left(\frac{V_\phi}{V} \right) > 0$
on the potential.

With the use of the slow-roll expressions for the scalar spectral
index $n_s=1-6 \epsilon + 2 \eta$ and for the tensor-to-scalar ratio $r=16 \epsilon$, 
 Eq. (\ref{inflaton}) can be rewritten as: 
\be 
\frac{d \langle Q^ {(1)\,2} \rangle}{d N} + \left( n_s - 1 + \frac{r}{8} \right) \langle
Q^ {(1)\,2} \rangle = \frac{H^2 (t)}{4 \pi^2} \,.
\label{inflaton2} 
\ee 
Eq. (\ref{inflaton2}) tell us that power-law inflation,  for which
$n_s - 1 = -r/8$ holds, lies at the threshold between two opposite
behaviors.  We note that Eq.
(\ref{inflaton2}) is the same for a modulus with the mass $m^2_\chi
= c H^2$ and $c = 3(n_s-1+r/8)/2$: below the power-law inflation
line Mukhanov variable behaves as a modulus with negative $c$.

The solution of Eq. (\ref{inflaton}) is
\be
\langle Q^ {(1)\,2}\rangle
= \frac{\epsilon(N)}{4 \pi^2} \int^N d n
\frac{H^2 (n)}{\epsilon (n)} \,,
\label{SolMukhanov}
\ee
where we have assumed  $\langle Q^{(1)\,2} \rangle (N_i)=0$.

%%%%%%%%%%%%%%%%%%%%%%%%%%%%%%%%%%%%%%%%%%%%%%%%%%%%%%%%%%%

\section{V. Inflationary models and results}

The detailed evolution of the expansion during the accelerated stage
depends on the inflaton potential and so does the growth of quantum fluctuations.
For this reason we consider in the following
four different potentials which are representative of the {\em ``inflationary zoo''}.
Since we shall study the growth of quantum fluctuations as a function 
of $N$,
the evolution of the Hubble parameter $H$ has to be expressed as a function of $N$, too.

The first obvious model is chaotic quadratic inflation, which we have also
used in our previous investigations \cite{FMVV_I,FMVV_II,FMVV_IV}:
\be
V(\phi) = \frac{m^2}{2} \phi^2  \,\,\,\,\,\,\,\,\,\,\,\,\,{\rm with}\,\,\,\,\,\,\,\,\,\,\,\,\,
H^2 \simeq H_i^2 - \frac{2}{3} m^2 N \label{htraj_quadratic} 
\,\,\,\,\,\,\,\,\,\,\,\,,\,\,\,\,\,\,\,\,
\dot \phi \simeq - \sqrt{\frac{2}{3}} m M_{\rm pl} \label{phitraj_quadratic}\,,
\ee
where $M_{pl}^{-2}=8\pi G$.
Let us note that in the numerical results presented in the figures
all dimensional quantities have been rescaled w.r.t.
$m_{pl}=M_{pl} \sqrt{8\pi}$. 

As another large field inflationary model we consider an exponential potential
\be
V = V_0 e^{-\frac{\lambda}{M_{\rm pl}} \phi}
\,\,\,\,\,\,\,\,\,\,\,\,\,{\rm with}\,\,\,\,\,\,\,\,\,\,\,\,\,
a(t)=\left(\frac{t}{t_i}\right)^p\,\,\,\,\,\,,\,\,\,\,\,\,H(t)=\frac{p}{t}
\ee
with $p = 2/\lambda^2$ \cite{exponential}.
The slow-roll conditions are well satisfied for $p >> 1$, and one obtains
\be
H = H_i \, e^{-N/p} = \frac{p}{t_i} \, e^{-N/p}
\,\,\,\,\,\,\,\,\,\,\,\,,\,\,\,\,\,\,\,\,\,\,\,\,
\phi = \sqrt{\frac{2}{p}} M_{pl} N + \phi_i \,.
\ee
We then consider the case of a
quadratic potential (of arbitrary sign) uplifted with an offset
$V_0$: 
\be 
V(\phi) = V_0 \pm \frac{M^2}{2} \phi^2 \,.
\label{hybridorsmall} 
\ee 
With the positive sign the potential in
Eq. (\ref{hybridorsmall}) is an approximation for the simplest
model of hybrid inflation well above the scale of the end of
inflation; in this case $\phi$ decreases during the inflationary
expansion. With the negative sign the potential in Eq.
(\ref{hybridorsmall}) is a simple small field inflation model,
again far from the end of inflation; in this case $\phi$ increases
during the inflationary expansion. In the following we use the
approximate solution for the square of the inflaton as: 
\be 
\phi^2(N) \simeq \phi_i^2 e^{ \mp 2 \frac{M^2 M_{\rm pl}^2}{V_0}
(N-N_i)} \,, 
\label{phiapprox} 
\ee 
valid when $H \simeq \sqrt{V_0/(3 M_{pl}^2)}$, i.e.
$\frac{M^2}{V_0}|\phi^2-\phi_i^2|\ll 1$.

Let us now compare the mean square of first order
quantum fluctuations of the test fields with a small effective
mass introduced above with that of the first order Mukhanov
variable.
For the inflationary zoo chosen the results are the following.

%%%%%%%%%%%%%%%%%%%%%%%%%%%%%%%%%%%%%%%%%%%%%%%%%%%%%%

\subsection{Chaotic quadratic inflation}
The growth of a test field with constant mass $m_\chi$ is described by \cite{us1,us2}
\be 
\langle \chi^{(1)\,2} \rangle
= \frac{3 H^{2 \frac{m^2_\chi}{m^2}}}{8
\pi^2 (2 m^2 - m_\chi^2)} ( H_i^{4-2 \frac{m^2_\chi}{m^2}}
- H^{4-2 \frac{m^2_\chi}{m^2}} )\,.
\ee
For the case $m_\chi^2=c H^2$, using Eq.(\ref{SolB}), we obtain \cite{us2}
\be 
\langle \chi^{(1)\,2}
\rangle = \frac{m^2}{6\pi^2}\Bigl[
\Bigl(1-e^{-\frac{2}{3} c N}\Bigr)\left(\frac{9}{4c^2}+
\frac{3}{2c}N_T\right)-\frac{3}{2c}N\Bigr]\,, 
\ee 
where $N_T=\frac{3H_i^2}{2m^2}$ is equal to maximal number of possible
e-folds in this chaotic model. In the limiting case $c\to 0$ and
at the end of inflation ($N=N_T-3/2$), we recover the
result\footnote{This corresponds to the massless
  limit of moduli production computed in  Eq. (15) of \cite{us1} for $\alpha \to 0$.}:
\be
\langle \chi^{(1)\,2} \rangle \simeq
\frac{m^2}{12\pi^2}N_T^2=\frac{3H_i^4}{16\pi^2m^2} \,.
\label{massless_moduli}
\ee
For a nonminimally coupled scalar field the integral in Eq.(\ref{SolC}) 
 can be
easily computed in a closed form in terms of the exponential
integral function $E_\nu(z)$ (see, for example, \cite{GR}). One finds
\begin{eqnarray}
\langle \chi^{(1)\,2} \rangle &\simeq&
\frac{m^2}{6\pi^2} \frac{e^{8 \xi (N_T-N)}}{(N_T-N)^{2\xi}}
\left[
(N_T-N)^{2+2\xi}E_{-1-2\xi}\bigl(8\xi(N_T-N)\bigr) \right. \nonumber \\
& & \left. 
-(N_T-N_i)^{2+2\xi}E_{-1-2\xi}\bigl(8\xi(N_T-N_i)\bigr)
\right]\,.
\end{eqnarray}
One can verify that in the limit $\xi \to 0$ at the end of
inflation and a fixed large value for $N_T$, the result of
Eq.(\ref{massless_moduli}) for a massless modulus is again
reobtained.
To conclude, for the Mukhanov variable Eq.(\ref{SolMukhanov}) gives \cite{FMVV_II,us1,us2}
\be
\langle Q^{(1)\,2} \rangle
= \frac{H_i^{6} - H^{6}}{8 \pi^2 m^2 H^2} \,.
\ee

The comparison between these results is showed in Fig.1.

%%%%%%%%%%%%%%%%%%%%%%%%%%%%%%%%%%%%%%%%%%%%%%%%%%%%%%%%%%%%%%%%%%%%%%%%%
\begin{figure}[h]
\centering
\includegraphics[height=.25\textwidth]{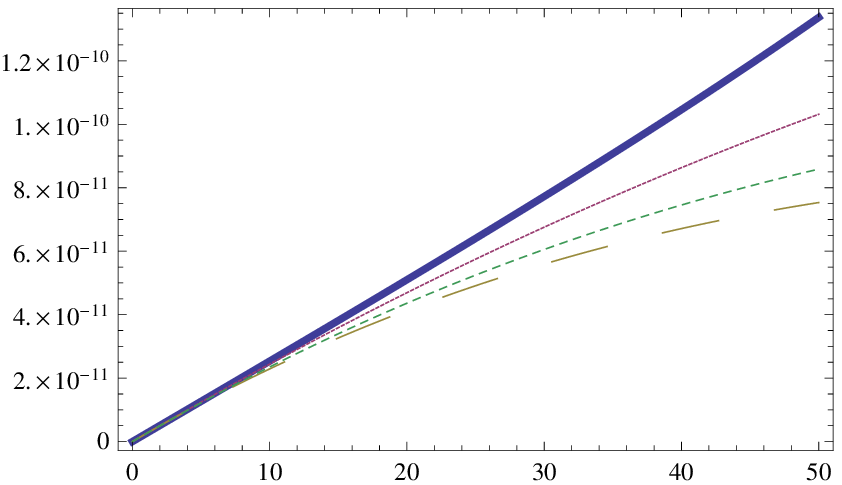}
\caption{\label{fig:quadratic}
Evolution of the mean square quantum fluctuations (in units of $m_{\rm pl}^2$)
versus the number of e-folds $N$ for the quadratic chaotic model.
We have chosen the inflationary trajectory 
with $m=10^{-6} \, m_{\rm pl}$ and $H_i = 10 \, m$. The mean square
Mukhanov variable (thick line) dominates over those of
test fields ($m_\chi \simeq 0.3 m$ is the solid line, $c=0.02$ is the dashed line, $\xi=0.001$ is the dotted line).}
\end{figure}

\subsection{Power-law inflation model}

The growth of a test field with constant mass $m_\chi$ is described by \cite{us1,us2}
\begin{eqnarray}
\langle \chi^{(1)\,2} \rangle &=& \frac{p}{8 \pi^2} H_i^2
\exp\left(-\frac{p}{3}\frac{m^2_\chi}{H^2}\right)
\left[-\exp\left(\frac{p}{3}\frac{m^2_\chi}{H^2}\right)\frac{H^2}{H_i^2}
\right. \nonumber \\ & & \left. 
+\frac{p}{3}\frac{m^2_\chi}{H_i^2}
Ei \left(\frac{p}{3}\frac{m^2_\chi}{H^2}\right)
+\exp\left(\frac{p}{3}\frac{m^2_\chi}{H_i^2}\right)-
\frac{p}{3}\frac{m^2_\chi}{H_i^2}
Ei \left(\frac{p}{3}\frac{m^2_\chi}{H_i^2}\right)\right] \,,
\end{eqnarray}
where $E_i$ is the exponential integral function (\cite{GR}).

For the case $m_\chi^2=c H^2$
\be
\langle \chi^{(1)\,2} \rangle = \frac{p}{8 \pi^2} H_i^2 \left(c
\frac{p}{3}
-1\right)^{-1}\left(e^{-2 \frac{N}{p}}-e^{-\frac{2}{3} c N}\right)\,. 
\ee
For a nonminimally coupled scalar field 
\be
\langle \chi^{(1)\,2} \rangle = \frac{p}{8 \pi^2} H_i^2 \left(-2
\xi-1
+4 p \xi\right)^{-1}\left(e^{-2 \frac{N}{p}}-e^{\xi N\left(\frac{4}{p}-8
\right)}\right) \,,
\ee
and, to conclude, for the Mukhanov variable, one obtains
\be
\langle Q^{(1)\,2} \rangle = \frac{p}{8 \pi^2}(H_i^2-H^2) \,,
\ee
which at late times (see also \cite{Marozzi}) becomes
$\langle Q^{(1)\,2} \rangle=\frac{p H_i^2}{8 \pi^2}$.

Fig. 2 presents comparison of results for the test fields 
and for the Mukhanov variable in the case of this second
inflationary model.

\begin{figure}[h!]
\centering
\includegraphics[height=.25\textwidth]{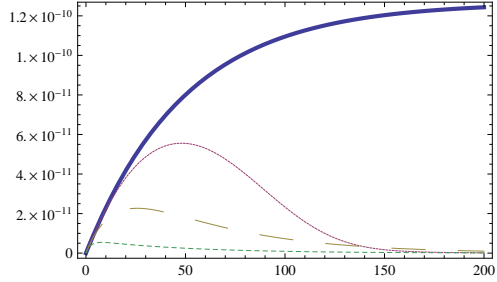}
\caption{\label{fig:powerlaw}
Evolution of the mean square quantum fluctuations (in units of $m_{\rm pl}^2$)
versus the number of e-folds $N$ for the exponential potential.
We have chosen the inflationary trajectory 
with $p=100$ and $t_i = 10^7 \, m_{\rm pl}^{-1}$. The mean square 
Mukhanov variable (thick line) dominates over those of
test fields ($m_\chi = 10^{-6} \, m_{\rm pl}$ is the solid line, $c=0.1$ is
the dashed line, $\xi=0.05$ is the dotted line).}
\end{figure}

\subsection{Small field inflation and Hybrid inflation models}
Let us consider the potentials in Eq. (\ref{hybridorsmall}), in the
lowest order approximation; that is, for $V \simeq V_0=3 H_0^2 M_{pl}^2$.

For the case of the growth of a test scalar field with constant mass we have
\be
\langle \chi^{(1)\,2} \rangle \simeq  \frac{3 H_0^4}{8 \pi^2 m_\chi^2}
\Bigl(1-e^{-\frac{2 m_\chi^2}{3 H_0^2}
N}\Bigr) \,.
\label{massivemod}
\ee

Let us note that  the  corrections induced by a non-zero
$M^2\phi^2 /V_0$ term are typically small both for the case of
hybrid inflation as well as for small field inflation (as long as
the field does not grow too much due to instability). 

For the case $m_\chi^2=c H^2$ we have 
\be
\langle \chi^{(1)\,2} \rangle = \frac{3 H_0^2}{8\pi^2 c}
\Bigl(1-e^{-\frac{2}{3} c N}\Bigr) \,,
\ee
while, for a nonminimally coupled scalar field one obtains
\be
\langle
\chi^{(1)\,2}\rangle = \frac{H_0^2}{32 \pi^2 \xi}
\Bigl(1-e^{-8 \xi N}\Bigr) \,. 
\ee
For the Mukhanov variable we obtain, 
by solving Eq.~(\ref{inflaton}) using the expression in
Eq.(\ref{phiapprox}) with no further approximations, the following result 
\be
\langle Q^{(1)\,2} \rangle \simeq
\pm \frac{4 V_0^2(1-y)+3M^4 \phi_i^2 y \left(4M_{pl}^2 (N-N_i) +
\phi_i^2(1-y)\right) \pm y(1-y^2)\frac{M^6}{4V_0}}
{96 \pi^2 M^2 M_{\rm pl}^4 \left(1\pm y \frac{M^2 \phi_i^2}{2V_0}\right)^2} \,,
\ee
where we have set $y=y(N)=e^{\mp\frac{2M^2 M_{\rm
pl}^2}{V_0}(N-N_i)}$. From this expression, when analyzing the
hybrid inflation case, one can notice that the fluctuations have a
maximum for a certain amount of e-folds $N$ and then decay to the
asymptotic value for a large N. 

Comparison of mean squares of fluctuations for these last two
models is presented in Figs. 3 and 4.
For the hybrid model, quantum fluctuations of test
fields with a small effective mass can dominate the
Mukhanov variable, because of the presence of the
leading constant term in the potential.

\begin{figure}[h]
\centering
\includegraphics[height=.25\textwidth]{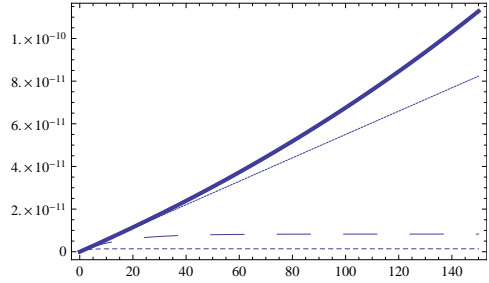}
\caption{\label{fig:small} Evolution of the mean square quantum
fluctuations (in units of $m_{\rm pl}^2$) versus the number of
e-folds $N$ for the small field inflationary model. 
For the inflationary background we have
chosen $V_0=2.6 \times 10^{-12} m_{\rm pl}^4$, $M = 0.85 \times
10^{-6} m_{\rm pl}$ and $\phi_i = 0.3 \, m_{\rm pl}$ as
parameters. The mean square Mukhanov variable
(thick line) dominates over those of test fields ($m_\chi =
10^{-2} H_0$ is the solid line, $c=0.1$ is the dashed line,
$\xi=0.05$ is the dotted line). }
\end{figure}

\begin{figure}[h!]
\centering
\includegraphics[height=.25\textwidth]{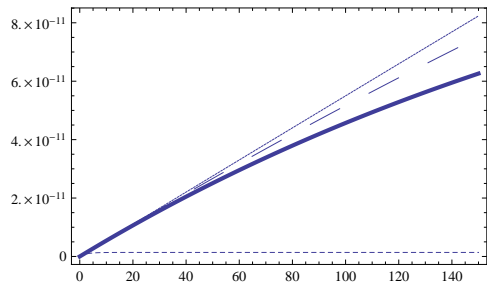}
\caption{\label{fig:hybrid} Evolution of the mean square quantum
fluctuations (in units of $m_{\rm pl}^2$) versus the number of
e-folds $N$ for the hybrid model. For
the inflationary background we have chosen $V_0=2.6 \times
10^{-12} m_{\rm pl}^4$, $M = 1.8 \times 10^{-6} m_{\rm pl}$ and
$\phi_i = 0.3 \, m_{\rm pl}$ as parameters. In this case the mean
square of moduli can dominate over the mean square of
Mukhanov variable (thick line): the parameters chosen
are $m_\chi = 10^{-2} H_0$ (solid line), $c=0.002$ (dashed line),
$\xi=0.05$ (dotted line). }
\end{figure}

\section{VI. Conclusions}
We have discussed the stochastic approach for general scalar
fluctuations, which may have a nonzero homogeneous mode, in a
generic model of inflation. We show, by using the field theory
results as a guideline, that the stochastic equations for the canonical
gauge invariant variable associated with such scalar fluctuations
are naturally formulated as a flow in terms of the number of
e-folds $N$.

We have then studied in detail the growth of quantum fluctuations
in realistic inflationary models with $\dot H \ne 0$. We have
selected four different potentials as representative examples of
the {\em inflationary zoo} and different types of nearly massless
fluctuations, including inflaton ones. 

We have found that for most of the inflationary models, the mean
square of the gauge invariant Mukhanov variable dominates over
those of moduli with a non-negative effective mass. Hybrid
inflationary models can be an exception: the mean square of a test
field can dominate over that of the gauge invariant Mukhanov 
variable on choosing parameters appropriately. Our findings
show that the understanding of inflaton dynamics including its
quantum fluctuations is more important than the moduli problem in
most of the inflationary models.

%%%%%%%%%%%%%%%%%%%%%%%%%%%%%%
{\bf Acknowledgements}
%%%%%%%%%%%%%%%%%%%%%%%%%%%%%%%%

G.M. wish to thank R. Durrer for stimulating discussions during the conference.

\end{document}